\documentclass{article}
\usepackage[T1]{fontenc}
\usepackage[utf8]{inputenc}
\usepackage{ismir} 
\usepackage{amsmath,cite,url}
\usepackage{graphicx}
\usepackage[table]{xcolor}
\usepackage{tabularx}
\usepackage{pifont}
\usepackage{booktabs}
\usepackage{titlesec}  
\usepackage{multirow}
\definecolor{cmarkgreen}{HTML}{1B7F3A}
\definecolor{xmarkred}{HTML}{B00020}
\newcommand{\cmark}{\textcolor{cmarkgreen}{\ding{51}}}
\newcommand{\xmark}{\textcolor{xmarkred}{\ding{55}}}
\newcommand{\ours}{SKY-Piano}

\title{SKY-Piano: A Multimodal Piano Performance Dataset}

\multauthor
  {\shortstack{%
    Joonhyung Bae$^{1}$ \hspace{0.5cm} Dawon Park$^{2}$ \hspace{0.5cm} Taegyun Kwon$^{1}$ \hspace{0.5cm} Yoon-Seok Choi$^{2}$ \hspace{0.5cm} Hyeon Hur$^{2}$\\[2pt]
    Satoshi Obata$^{3}$ \hspace{0.5cm} Shigeru Kai$^{3}$ \hspace{0.5cm} Yohei Wada$^{3}$ \hspace{0.5cm} Yu Takahashi$^{3}$ \hspace{0.5cm} Akira Maezawa$^{3}$\\[2pt]
    Jaebum Park$^{2}$ \hspace{0.5cm} Jonghwa Park$^{2}$ \hspace{0.5cm} Juhan Nam$^{1}$%
  }}
  {$^{1}$ KAIST \qquad $^{2}$ Seoul National University \qquad $^{3}$ Yamaha Corporation\\[2pt]
   {\tt\small \{jh.bae, juhan.nam\}@kaist.ac.kr, \{dw\_park, dallas71, sunny000927, parkpe95\}@snu.ac.kr}\\
   {\tt\small \{satoshi.obata, shigeru.kai, yohei.wada, yu.takahashi, akira.maezawa\}@music.yamaha.com}\\
   {\tt\small \{ilcobo2, pianistpark\}@gmail.com}
  }

\def\authorname{J. Bae, D. Park, T. Kwon, Y.-S. Choi, H. Hur, S. Obata, S. Kai, Y. Wada, Y. Takahashi, A. Maezawa, J. Park, J. Park, and J. Nam}

\begin{document}

\maketitle

\begin{abstract}
Music information retrieval research on piano performance increasingly involves diverse modalities of data and annotations beyond audio and MIDI. 
We present \textbf{\ours{}},\footnote{Project page: \url{https://joonhyungbae.github.io/skypiano/}, with the interactive explorer, the pipeline code, and the dataset. The dataset name is an acronym of the contributing institutions (Seoul National University, KAIST, and Yamaha).} a multimodal piano performance dataset that includes  11~hours of performance recordings of motion, multi-view video, audio, MIDI from 7 professional and 12 amateur pianists along with MusicXML scores. The performance pieces were selected considering playing technique, difficulty, and performer expertise on a shared core repertoire.
The motion data include both hand and body motion, released in both flagged form, where samples lost to marker occlusion are marked as unreliable, and imputed form, where those gaps are reconstructed, together with Visual3D body-segment kinematics and other time-synchronized modalities.
To easily browse different modalities of data at a glance, we provide an interactive web browser.  In addition, we developed a fingering annotation model and tool for deriving pseudo fingering annotations from the MIDI and motion data. 
Lastly, we present MIDI-to-motion generation through a fine-tuning experiment as a use case of the dataset. 
\end{abstract}

\begin{table*}[t]
\centering
\caption{Comparison of multimodal piano performance datasets. \cmark\ available, \xmark\ not available, - not applicable. ``Score'' denotes composer-intended symbolic notation aligned to the performance, distinct from performance MIDI; ``unlabeled'' marks web-collected data with no per-pianist skill classification.}
\label{tab:comparison}
\tiny
\setlength{\tabcolsep}{3pt}
\renewcommand{\arraystretch}{1.05}

\resizebox{\textwidth}{!}{%
\begin{tabular}{l|cc|cc|c|c|c|c|c|c|c|c}
\toprule
\multirow{2}{*}{Dataset} & \multicolumn{2}{c|}{Hand} & \multicolumn{2}{c|}{Body} & \multirow{2}{*}{Video} & \multirow{2}{*}{Audio} & \multirow{2}{*}{MIDI} & \multirow{2}{*}{Score} & \multirow{2}{*}{Fingering} & Pianists & \multirow{2}{*}{Pieces} & \multirow{2}{*}{Hours} \\
& Real & Pseudo & Real & Pseudo & & & & & & Professional/Amateur & & \\
\midrule
DAEMG~\cite{sarasua2017datasets} & \xmark & \xmark & \cmark & - & 1-view & \cmark & \cmark & \xmark & \xmark & 2/- & 1 & 7 \\
Solos~\cite{montesinos2020solos} & \xmark & \xmark & \xmark & \cmark & 1-view & \cmark & \xmark & \xmark & \xmark & unlabeled & - & 12 \\
MOSA~\cite{huang2024mosa} & \xmark & \xmark & \cmark & - & \xmark & \cmark & \xmark & \xmark & \xmark & 8/- & 10 & 15 \\
F\"{u}r Elise~\cite{wang2024furelise} & \xmark & \cmark & \xmark & \xmark & \xmark & \cmark & \cmark & \xmark & \xmark & 15/- & 153 & 10 \\
PianoMotion10M~\cite{gan2024pianomotion10m} & \xmark & \cmark & \xmark & \xmark & 1-view & \cmark & \xmark & \xmark & \xmark & unlabeled & - & 116 \\
PianoKPM~\cite{liu2025pianoemg} & \xmark & \cmark & \xmark & \xmark & \xmark & \xmark & \xmark & \xmark & \xmark & 20/- & 7 & 12.6 \\
PianoVAM~\cite{kim2025pianovam} & \xmark & \cmark & \xmark & \xmark & 1-view & \cmark & \cmark & \xmark & \cmark & -/10 & 106 & 21.0 \\
\midrule
\rowcolor{gray!20}
\textbf{Ours (\ours{})} & \cmark & - & \cmark & - & up to 4-view & \cmark & \cmark & \cmark & \cmark & 7/12 & 35 & 11.0 \\
\bottomrule
\end{tabular}%
}
\end{table*}

\section{Introduction}\label{sec:introduction}

A piano performance is shaped not only by which notes are played, but by how they are played. Hand posture, weight transfer, and the wrist rotation between phrases all shape the sound, yet none of them appear in audio or MIDI alone. Recent MIR research therefore couples audio and MIDI with hand and body motion to support tasks such as motion-conditioned hand-pose generation~\cite{bae2026tipianocascadedpianohand,wang2024furelise,gan2024pianomotion10m,liu2025separate}, audio-visual piano transcription~\cite{koepke2020sight,li2023crnngcn,li2024twostage,kim2025pianovam}, expressive-performance modeling~\cite{park2023multivariate,goebl2013temporal}, and fingering estimation~\cite{nakamura2020statistical,ramoneda2022autoregressive}.

The challenge is precision. Distinguishing which finger pressed which key requires sub-centimeter motion data, and small inaccuracies blur the articulation that separates one performer from another. Three requirements emerge. First, the motion must be directly measured rather than estimated. Second, the streams must be frame-synchronized across audio, MIDI, score, and video. Lastly, the repertoire must let researchers compare performers on shared material under matched conditions.

However, no existing corpus satisfies all three at once. Pseudo-motion approaches that estimate hand motion from video~\cite{gan2024pianomotion10m,wang2024furelise} cover large repertoires but incur per-joint errors of several centimeters, blurring exactly the keystroke-level distinctions that fine-grained MIR analyses depend on. The two real-mocap piano corpora are limited in different ways. DAEMG~\cite{sarasua2017datasets} contains only one piece played by two professionals, and MOSA~\cite{huang2024mosa} omits hand motion and amateurs despite a richer ten-piece repertoire. No prior dataset combines technique, difficulty, and skill variation on measured motion.

We present \ours{}, an 11-hour multimodal piano performance dataset of 19 pianists (7 professional, 12 amateur) that closes this gap with three contributions. First, we release directly measured hand and body motion from professional optical mocap, frame-synchronized with audio, MIDI, multi-view video, MusicXML scores, and Visual3D body-segment kinematics. Second, the repertoire is structured to cross two technique categories (movement-oriented exercises, and standard pianistic patterns such as scales, arpeggios, and octave passages) with three difficulty levels and two expertise tiers on a shared core, enabling controlled comparison across performers and musical structure within one corpus. Third, we provide a mocap-grounded pseudo fingering pipeline, an interactive web browser for navigating the synchronized streams, and a Tipiano~\cite{bae2026tipianocascadedpianohand} fine-tuning baseline as a representative use case. \ours{} brings precise multimodal piano performance capture together with controlled musical structure in a single corpus.

\begin{figure*}[t]
\centering
\includegraphics[width=0.92\textwidth]{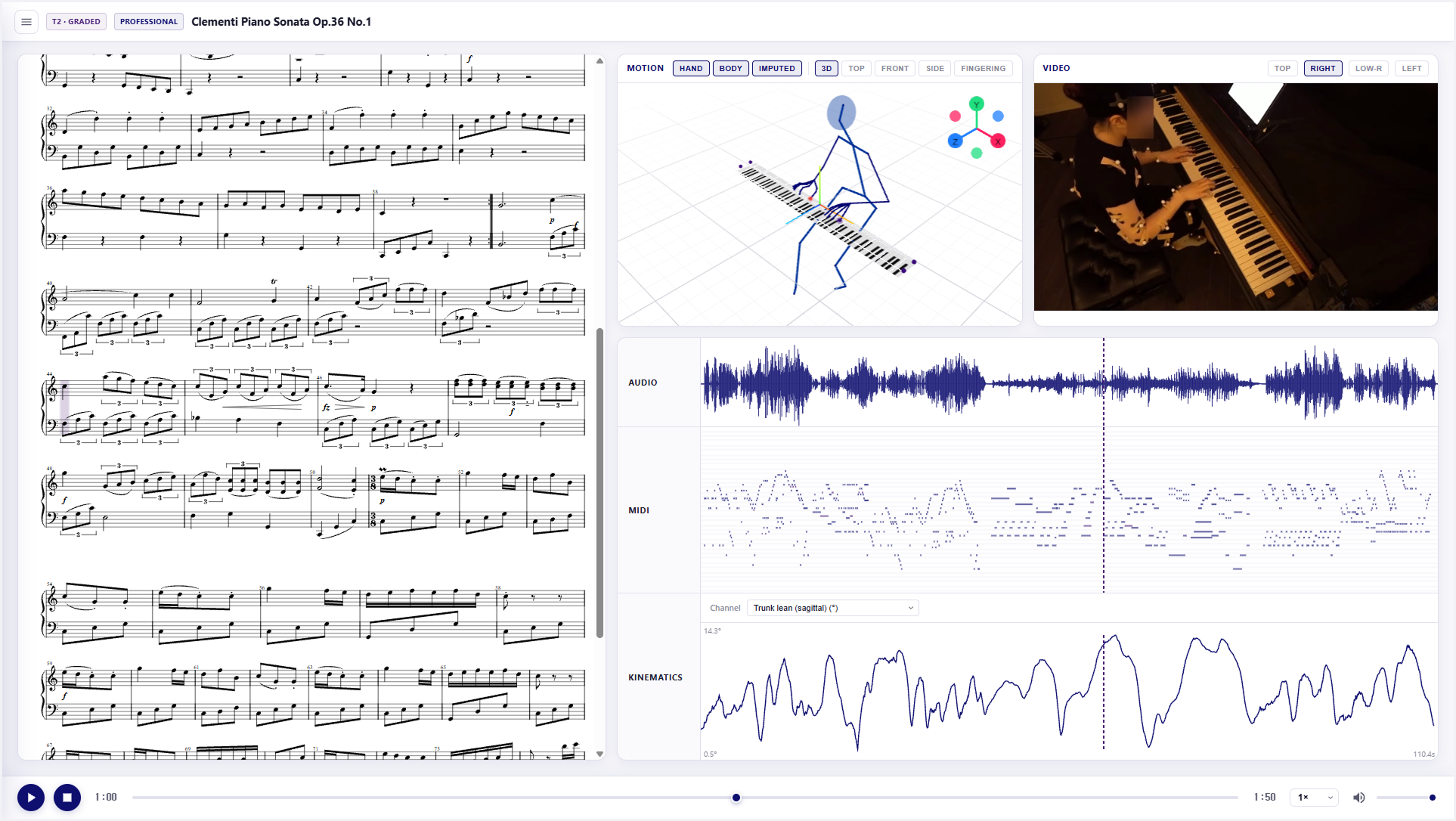}
\caption{\ours{} at a glance. Hardware-synchronized modalities for one released trial in the interactive viewer.}
\label{fig:teaser}
\end{figure*}

\section{Related Work}\label{sec:related}

\subsection{Multimodal Piano Performance Datasets}

\tabref{tab:comparison} summarizes existing datasets. Only two carry real motion capture: DAEMG~\cite{sarasua2017datasets} provides body motion for a single piece, and MOSA~\cite{huang2024mosa} adds rich metadata across 10 pieces but omits MIDI as well as the hand motion most informative for piano-specific MIR. The rest rely on motion estimated from video. Solos~\cite{montesinos2020solos} and PianoMotion10M~\cite{gan2024pianomotion10m} reach large scale this way; F\"{u}r Elise~\cite{wang2024furelise} pairs pseudo hand motion with MIDI across 153 pieces; PianoKPM~\cite{liu2025pianoemg} adds surface EMG but omits audio and MIDI; and PianoVAM~\cite{kim2025pianovam} pairs audio, MIDI, and top-view video with pseudo hand landmarks and pseudo fingering, without body data. None of them combine real hand and body motion with the simultaneous factorization of playing technique, difficulty, and performer expertise that \ours{} provides.

\subsection{Piano Performance Movement Analysis}

Goebl and Palmer~\cite{goebl2013temporal} found that skilled pianists exhibit more efficient hand movement. Park et al.~\cite{park2023multivariate} applied PCA to piano performance kinematics and identified the thumb, little finger, and wrist as dominant axes across techniques, though within a single technique category. \ours{} extends such analyses across technique, difficulty, and expertise on a shared repertoire.

\subsection{MIDI-to-Motion Generation}

Hand motion generation from MIDI for piano divides broadly into physics-based and kinematic-based approaches. Physics-based methods train reinforcement-learning policies on simulated pianos~\cite{robopianist2023,qian2024pianomime,huang2025pandoradiffusionpolicylearning}, prioritizing precise key contact under physical constraints. Kinematic-based methods predict joint trajectories directly from motion-capture corpora, including diffusion-based generation~\cite{gan2024pianomotion10m,liu2025separate} and cascaded skeleton synthesis~\cite{bae2026tipianocascadedpianohand}. As a representative kinematic-based system, Tipiano~\cite{bae2026tipianocascadedpianohand}, trained on the F\"{u}r Elise~\cite{wang2024furelise} pseudo-motion corpus, predicts hand motion from MIDI plus fingering as the 21 keypoints (wrist plus four joints per finger) of MANO~\cite{romero2017mano}, a parametric 3D hand model widely used for hand-pose estimation, reporting a key-contact F1 of 0.910 and a mean per-joint position error (MPJPE) of 32.8~mm on its held-out test set. Kinematic-based metrics are bounded by the underlying motion fidelity (centimeter-scale for video-derived corpora), motivating measured-motion datasets.

\subsection{Fingering Estimation}

Piano fingering estimation has been studied via hidden Markov models~\cite{nakamura2020statistical} and autoregressive neural networks~\cite{ramoneda2022autoregressive} from symbolic scores, and via a geometry-based candidate-scoring algorithm over synchronized video and MIDI in PianoVAM~\cite{kim2025pianovam}. Our fingering pipeline (\secref{sec:fingering}) adapts PianoVAM's methodology to real motion capture, replacing video-based $z$-depth estimation with directly measured fingertip $z$ and adding a $z$-onset refinement that lifts coverage and precision while preserving PianoVAM's explicit-ambiguity reporting.

\section{Dataset Acquisition and Pre-processing}\label{sec:dataset}

\figref{fig:teaser} shows the interactive web explorer surfacing score, hand/body motion, top-view video, audio, MIDI, and Visual3D kinematics on a common timeline. Pseudo fingering is rendered as a per-frame color overlay on the pressing fingertip in the hand-motion view.

\subsection{Piece Selection}\label{sec:repertoire}

The structured portion of \ours{} contains 26 controlled task items, comprising six Technique~1 exercises, five Technique~2 exercises, and 15 graded pieces spanning beginner to advanced levels (\tabref{tab:repertoire}); the release additionally includes professional free-piece titles. The technique exercises cross performer expertise and the graded pieces cross difficulty within the professional tier, so together they cross technique, difficulty, and expertise on a shared core, whereas the free pieces expose more idiosyncratic high-level performance behavior. Both were selected collaboratively by three expert pianists (one with 19 years of pedagogy experience) and the research team.

The two technique categories are drawn from Fink's ``Mastering Piano Technique''~\cite{fink1992mastering}, with eleven exercises selected for kinematic distinctiveness. Graded pieces span three difficulty levels anchored to the Royal Conservatory of Music Piano Syllabus~\cite{rcm2015piano}. Most graded pieces were recorded as excerpts rather than complete works; the released MusicXML for each piece is trimmed to the recorded passage, so the score alignment (\secref{sec:preproc}) states the covered extent.

\begin{table}[t]
\centering
\caption{Selected repertoire.}
\label{tab:repertoire}
\scriptsize
\setlength{\tabcolsep}{4pt}
\renewcommand{\arraystretch}{1.05}
\setlength{\tabcolsep}{3pt}
\begin{tabularx}{\columnwidth}{@{}l@{\hspace{6pt}}X@{}}
\toprule
\textbf{Category} & \textbf{Items} \\
\midrule
\textbf{Technique 1} & Adduction/abduction; making short and long sounds; playing with strength in the upper arm; finger extension/flexion; press forward with force--rebound; bounce back hand. \\
\textbf{Technique 2} & Scale, arpeggio, octave repetition, octave scale, broken octave arpeggio. \\
\midrule
\multicolumn{2}{@{}l}{\textbf{Graded pieces}} \\
\quad Beginner & Bach Minuet in G Major (BWV Anh.~114); Clementi Piano Sonata Op.~36 No.~1. \\
\quad Intermediate & Chopin Prelude in E Minor Op.~28 No.~4; Beethoven \emph{Moonlight Sonata} (1st mvt); Mozart Sonata No.~16 K.~545; Schumann \emph{Tr\"{a}umerei} Op.~15 No.~7; Bach Prelude No.~1 in C Major (WTC~I); Mozart 12 Variations on ``Twinkle Twinkle Little Star'' K.~265. \\
\quad Advanced & Chopin Waltz in C\# Minor Op.~64 No.~2; Chopin Waltz in E-flat Major Op.~18; Chopin \emph{Minute Waltz} Op.~64 No.~1; \emph{Silvery Waves}; Schubert Impromptu Op.~90 No.~4; Chopin Prelude in D-flat Major Op.~28 No.~15 (``Raindrop''); Chopin Nocturne in E-flat Major Op.~9 No.~2. \\
\midrule
\textbf{Free (pro)} & Chopin Etude Op.~10 No.~4; Chopin Etude Op.~10 No.~10; Chopin Mazurka Op.~17 No.~4; Chopin Sonata No.~2 (1st mvt); Bach Prelude and Fugue in C\# Major (BWV 848); Ravel \emph{La Valse}; Ravel \emph{Jeux d'eau}; Ravel \emph{Gaspard de la Nuit} (Ondine); Scriabin Sonata No.~2 (both mvts). \\
\bottomrule
\end{tabularx}
\end{table}

In addition to the structured repertoire, five professional pianists each contributed approximately 11~minutes of free-repertoire performance from the titles listed under \emph{Free (pro)} in \tabref{tab:repertoire}. One amateur pianist also contributed a separate free-repertoire session of similar duration. Free-repertoire recordings ship at the take level (audio, MIDI, and motion). The piece-split fingering subset is smaller than the recorded volume.

\subsection{Acquisition}

All released data was cleared for public release under per-modality consent (\secref{sec:ethics}). The dataset comprises 7 professional and 12 amateur pianists, with one amateur shared between the Amateur and Mixed cohorts so the 20 cohort entries in \tabref{tab:modality} deduplicate to 19 unique participants. The professional cohort anchors the modality-complete subset (audio, MIDI, hand/body motion, Visual3D (C-Motion, Inc., Germantown, MD, USA) body-segment kinematics, pseudo fingering, and four-view face-anonymized video). Professionals have conservatory training and active performance careers. Amateurs have basic to intermediate skills.

\begin{table}[t]
\centering
\caption{Per-cohort subject counts (professional / amateur) and released modality availability. \cmark\ (present), \xmark\ (absent).}
\label{tab:modality}
\scriptsize
\setlength{\tabcolsep}{4pt}
\renewcommand{\arraystretch}{1.05}
\begin{tabular*}{\columnwidth}{@{\extracolsep{\fill}}l c c c c c c c c}
\toprule
\textbf{Cohort} & \textbf{Pro/Am} & \textbf{Audio} & \textbf{MIDI} & \textbf{Hand} & \textbf{Body} & \textbf{Video} & \textbf{V3D} & \textbf{Fing.} \\
\midrule
Professional      & 5 / 0  & \cmark & \cmark & \cmark & \cmark & 4-view  & \cmark & \cmark \\
Amateur           & 0 / 11 & \cmark & \cmark & \cmark & \cmark & \xmark  & \cmark & \cmark \\
Mixed             & 2 / 2  & \cmark & \cmark & \cmark & \cmark & \xmark  & \cmark & \cmark \\
\bottomrule
\end{tabular*}
\end{table}

\begin{figure}[t]
\centering
\includegraphics[width=\columnwidth]{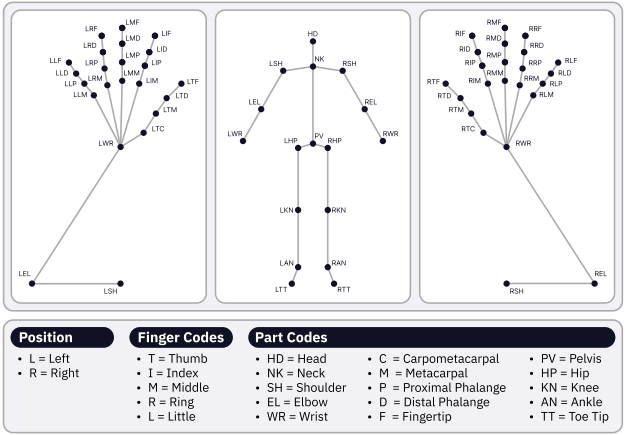}
\caption{Marker and joint layout. The left and right panels show per-side OptiTrack marker placement (23 markers per side, comprising 4 markers per finger, a wrist anchor, and elbow and shoulder anchors for arm tracking), totaling 46 markers across both arms. The center panel shows the 17-joint Visual3D-derived body-joint layout. The legend defines position (L/R), finger codes, and part codes.}
\label{fig:marker-layout}
\end{figure}

\tabref{tab:capture} summarizes the capture equipment. Body mocap follows the optical-marker convention of prior piano releases such as MOSA~\cite{huang2024mosa}, paired with synchronous OptiTrack hand mocap to capture posture and fingertip articulation. The hand marker layout is compatible with widely-used hand models (MANO, MediaPipe Hands), enabling direct mapping to motion-from-MIDI and hand-pose pipelines. \figref{fig:marker-layout} shows the per-hand marker layout, Visual3D body-joint layout, and part-code abbreviations. Four synchronized cameras capture the performance from top, left, right, and low-right angles, providing multi-view coverage of the keyboard, hands, and upper body.

Later sessions were recorded under a hardware timecode setup: a Rosendahl nanoSync generator driven by a Black Magic master clock stamps every stream with a common 30~fps SMPTE (Society of Motion Picture and Television Engineers) timecode, the broadcast standard for labeling each frame with a wall-clock time, so all streams share one frame index. The earlier sessions predate this setup and are aligned post hoc (\secref{sec:preproc}). In both cases, Disklavier MIDI is recorded through a digital audio workstation that timestamps it against the same SMPTE clock as the audio. To safeguard against software-induced drift, we apply a per-trial CQT-based audio-to-MIDI alignment from MAESTRO~\cite{hawthorne2019maestro} ($\sim$3~ms accuracy) via PianoVAM's~\cite{kim2025pianovam} Python reimplementation.

\begin{table}[t]
\centering
\caption{Capture equipment summary.}
\label{tab:capture}
\scriptsize
\setlength{\tabcolsep}{4pt}
\renewcommand{\arraystretch}{1.05}
\newcolumntype{Y}{>{\hsize=.85\hsize\raggedright\arraybackslash}X}
\newcolumntype{Z}{>{\hsize=1.15\hsize\raggedright\arraybackslash}X}
\begin{tabularx}{\columnwidth}{@{}l@{\hspace{6pt}}Y@{\hspace{6pt}}Z@{}}
\toprule
\textbf{Stream} & \textbf{System} & \textbf{Spec} \\
\midrule
Hand mocap & OptiTrack, 8 IR cams & 120~Hz, 23 reflective markers per side (4~mm) including elbow and shoulder anchors \\
Body mocap & Qualisys, 8 IR cams & 120~Hz, full-body marker suit, 17-joint Visual3D model \\
Audio & Focusrite Scarlett & 48~kHz / 24-bit WAV \\
MIDI & Yamaha Disklavier DC7X ENPRO & direct from instrument (velocity, timing, pedals) \\
Video & 4 synced cameras & 4K / 60~fps, top/left/right/low-right, HD downsampled, face-anonymized on released views (\secref{sec:anon}) \\
Sync & Rosendahl + Blackmagic master & 30~fps SMPTE to eSync2 / Qualisys / HDMI / audio \\
\bottomrule
\end{tabularx}
\end{table}

\subsection{Data Preprocessing}\label{sec:preproc}

\begin{figure}[t]
\centering
\includegraphics[width=0.88\columnwidth]{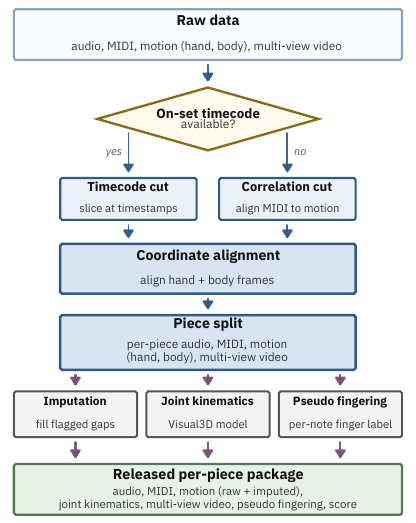}
\caption{Release-facing preprocessing pipeline.}
\label{fig:pipeline}
\end{figure}

\figref{fig:pipeline} shows the release-facing preprocessing flow as a common spine (trialization, coordinate alignment, piece splitting and identification) followed by a piece-level fan-out into modality-specific branches. Every preprocessing stage closes with a manual review pass before public release.

\paragraph{Trialization.}
Raw session recordings are first segmented into individual performance takes, or trials. For timecode-synchronized sessions, take boundaries are read directly from the markers the recording operator inserted into the timecode stream. For the earlier sessions without that setup, boundaries are recovered from whatever signal carries them (filename, motion, or audio) and then verified by hand, trial by trial.

\paragraph{Coordinate alignment.}\label{sec:align}
For timecode-synchronized sessions, hand and body streams inherit the shared timecode and align directly. For the earlier sessions, the streams are first temporally synchronized via cross-correlation of wrist velocities and verified per trial. Streams are then spatially aligned via weighted orthogonal Procrustes (Horn's method) over six bilateral landmarks (shoulders, elbows, and wrists) weighted by quasi-rigidity, expressing body motion in the hand-capture coordinate frame with median per-trial landmark residuals at the centimeter scale. For inter-session alignment, four stationary reference markers at the keyboard ends, visible in \figref{fig:teaser}, define a consistent axis across sessions.

\paragraph{Piece splitting and identification.}
Trialized and aligned streams are split into per-piece segments by MIDI silence-gap detection. Each piece's identity is determined by a symbolic-MIDI fingerprint matcher (pitch-class + IOI n-gram hashing in the spirit of constellation fingerprints~\cite{wang2003shazam}, restricted to a task-appropriate anchor library) and cross-checked manually against the recording log before release.

\paragraph{Per-modality post-processing.}
Released video\label{sec:anon} undergoes face anonymization with EgoBlur~\cite{raina2023egoblur} on the four views of the professional cohort, the only group with per-participant video consent (\tabref{tab:modality}). Hand and body motion are shipped in both flagged form (raw samples with validity flags) and imputed form using Self-Attention-based Imputation for Time Series (SAITS)~\cite{du2023saits}. Pseudo fingering (\secref{sec:fingering}) uses the flagged hand motion and piece MIDI. Per-piece MusicXML scores are note-aligned to the performance MIDI using Parangonar's automatic note matcher~\cite{peter2023nasap} and verified by a musicologist on every released piece.

\section{Dataset Statistics}\label{sec:stats}

The release totals 11~hours of synchronized recording across 19 unique pianists, averaging 39.4~min per professional (range 30--53, std 6.4) and 32.1~min per amateur (range 20--56, std 11.3). \tabref{tab:cohort-task} breaks the release down by group and task.

\begin{table}[t]
\centering
\caption{Release breakdown by group and task. Each cell shows participants covering that task / total MIDI notes. \xmark\ marks cells the group did not perform by design.}
\label{tab:cohort-task}
\scriptsize
\setlength{\tabcolsep}{4pt}
\renewcommand{\arraystretch}{1.05}
\begin{tabular*}{\columnwidth}{@{\extracolsep{\fill}}l c c c c}
\toprule
\textbf{Group} & \textbf{Tech.~1} & \textbf{Tech.~2} & \textbf{Set~1 graded} & \textbf{Free} \\
\midrule
Professional (7)   & 7 / 47{,}827 & 7 / 34{,}703 & 5 / 24{,}292 & 1 / 6{,}201 \\
Amateur (12)       & 12 / 81{,}989 & 12 / 25{,}875 & \xmark         & 1 / 5{,}613 \\
\midrule
\textbf{Total}     & 19 / 129{,}816 & 19 / 60{,}578 & 5 / 24{,}292 & 2 / 11{,}814 \\
\bottomrule
\end{tabular*}
\end{table}

Technique~1 and Technique~2 carry the controlled-comparison axes spanning both groups. Every Technique~2 sitting comprises the same five drill exercises of \tabref{tab:repertoire} at two target tempi each (slow and fast), so the C-major slow scale alone is performed by all 19 unique participants (7 professional, 12 amateur). Technique~1 anchors a second cross-group comparison over the same Fink~\cite{fink1992mastering} exercise book, also at 19 of 19. Set~1 graded pieces are professional-only by recording design, and free repertoire is performer-selected. The piece-split fingering subset covers one professional and one amateur sitting, while the take-level recordings are broader (\secref{sec:repertoire}).

Marker occlusion concentrates in distal finger landmarks. Across all 565 piece-level hand CSVs and 46 marker channels, per-marker mean missing-cell rate ranges from 6\% on wrist and proximal-finger markers to 44\% on the thumb-carpal pair, with a corpus-median of 14\%. These dropout patterns are consistent with prior pianist-marker preprocessing studies~\cite{kwon2023automated}. Releasing the motion in both forms (\secref{sec:preproc}) lets users choose between the conservative stream, in which occluded samples stay marked as missing, and the reconstructed stream, in which those gaps are filled by the imputation model, depending on whether their analysis can tolerate synthetic samples.

\section{Annotation of Fingering Labels}\label{sec:fingering}

\begin{figure}[t]
\centering
\includegraphics[width=0.88\columnwidth]{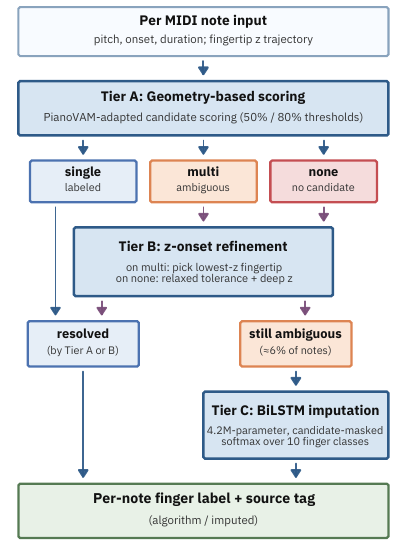}
\caption{Fingering annotation pipeline.}
\label{fig:fingering}
\end{figure}

We generate pseudo fingering labels by adapting PianoVAM's~\cite{kim2025pianovam} geometry-based scoring to motion capture. The idea is simple: at the moment a MIDI note sounds, the finger that pressed it is the one whose measured fingertip sits inside that key and lowest in depth. The pipeline therefore proceeds in three tiers, from the notes where that evidence is unambiguous to the notes where it is not. Because the fingertip $z$ is now measured rather than estimated from video, the joint scoring factors into a \emph{keystroke detection} stage (which key is pressed at which time) followed by a \emph{finger assignment} stage (which finger pressed it), and the explicit-ambiguity reporting from PianoVAM (single, multi-candidate, no-candidate) carries over without modification. \figref{fig:fingering} summarizes the resulting three-tier flow. Tier~A applies PianoVAM-derived geometric scoring. Tier~B refines Tier~A's multi-candidate and no-candidate cases via $z$-onset evidence. Tier~C imputes any residual ambiguity with a BiLSTM.

A keyboard coordinate frame with along-key axis $u$, across-key axis $w$, and depth axis $z$ is anchored on the two end-of-keyboard reference markers, with each key mapped to a 3D rectangle using the Yamaha piano's fixed key geometry. For each MIDI note, every fingertip accumulates a duration-averaged score ($1.0$ inside the key region, quadratic falloff within a half-key tolerance), only on frames where $z$ is below the press threshold. Tier B operates in a $\pm 25$~ms onset window. The multi-candidate pass requires a $\geq 5$~mm $z$ gap between the lowest fingertip and the runner-up. The no-candidate pass relaxes $(u, w)$ tolerance but enforces strict press-depth, rescuing only unambiguous $z$ dips. Residual ambiguity is retained as an explicit flag, and we release event-level fingering labels and frame-level contact labels.

The keystroke-detection stage admits fully automatic validation against MIDI ground truth, unlike PianoVAM's video-only setting. Across all 113{,}023 MIDI notes in the professional subset, a fingertip enters the target key region within $\pm 50$~ms of the MIDI onset and below 15~mm above the key surface for 91.4\% of notes (range 86--94\% across tasks). The median fingertip $z$ at the registered press is $-10.1$~mm relative to the key surface, confirming actual key depression rather than mere proximity.

For finger-assignment validation, a music technology researcher hand-labeled every MIDI note in three Set1 graded pieces (Beginner Clementi Op.~36 No.~1, Intermediate Mozart K.~545, and Advanced Schubert Op.~90 No.~4), each performed by a different professional pianist, totaling 2{,}269 notes. The audited trials are the exact trials shipped with the release, so the evaluation is reproducible from public materials. Annotation used a custom interface (included in the code release) over synchronized overhead video and 3D hand-marker visualization. On this audit, the geometry-based labeler with $z$-onset refinement reaches strict precision of 94.5\% over its 94.0\%-coverage committed subset. PianoVAM~\cite{kim2025pianovam} reports 95.6\% strict precision on a 1{,}500-note video-only audit but over only $\sim$82\% of notes, leaving 18\% without an imputation tier. \ours{}'s Tier C closes this gap to 100\% coverage.

Tier C is trained in the masked-fill spirit of the SAITS~\cite{du2023saits} marker imputation (\secref{sec:preproc}). Per note it consumes pitch, inter-onset interval, duration, and either the geometry-derived finger label or a mask token. For multi-candidate notes, the softmax is masked to the algorithm's candidate set, ensuring a geometrically plausible prediction. We train on the modality-complete professional subset (excluding the Mixed cohort), augmented with PIG~\cite{nakamura2020statistical} as a symbolic prior, and recover the algorithm labels at 84.3\% on a 10\% trial-level holdout (random-chance 10\% over 10 classes). Each released note carries a source tag (algorithm, imputed, or null).

\section{Applications}\label{sec:example}

\subsection{Supported MIR Tasks}

The synchronized audio and MIDI tracks enable audio-only piano transcription~\cite{hawthorne2019maestro}, and the multi-view video extends this to visual and audio-visual transcription~\cite{koepke2020sight,li2024twostage,kim2025pianovam}. MIDI paired with the released fingering labels enables automatic fingering estimation~\cite{nakamura2020statistical,ramoneda2022autoregressive}. The shared core repertoire crossed with the professional/amateur cohort design supports systematic skill-tier analysis and expressive-performance modeling~\cite{goebl2013temporal,park2023multivariate}. The hand and body mocap streams add motion-conditioned generation, biomechanical analysis of piano technique, and motion-augmented transcription.

\subsection{Case Study: Hand Motion Generation}

As one concrete instantiation we formulate \emph{hand motion generation from MIDI and fingering}. Inputs are MIDI events (pitch, onset, offset, velocity, pedal) and per-note pseudo fingering (\secref{sec:fingering}), and the target is the 120~Hz hand-marker trajectory in the keyboard frame of \secref{sec:align}. Two evaluation protocols follow from the release structure: \emph{leave-one-pianist-out}, which uses the manifest's stable per-participant identifier to keep folds leak-free across sessions, and \emph{leave-one-piece-out}, which holds out one piece per technique/difficulty stratum. Suggested metrics are key-press precision/recall, per-joint position error against the imputed mocap, per-finger jerk, and inter-finger collision rate.

We fine-tune Tipiano~\cite{bae2026tipianocascadedpianohand} on the Professional cohort under 5-fold leave-one-pianist-out, with the held-out pianist's pieces as the test set. MPJPE is measured across all 21 hand joints, in millimeters. Key-contact F1 registers a press when a fingertip's $z$ depth crosses thresholds calibrated against the Disklavier's acoustic trigger depth, following Tipiano's protocol. \tabref{tab:tipiano-loso} reports the comparison against Tipiano's published in-domain numbers on Für Elise~\cite{wang2024furelise}. Fine-tuning reduces MPJPE by 25.9\%, bringing the cross-domain error within 1.5$\times$ of Tipiano's in-domain MPJPE, while key-contact F1 moves in the opposite direction. That opposition traces to a fingertip-definition mismatch: Tipiano targets MANO joints, whereas \ours{} fingertips are physical markers on the fingernail, displaced several millimeters from the joint. Zero-shot predictions sit in the joint convention, matching the detector, and fine-tuning shifts them onto the marker convention, matching the markers but exiting the detector's calibrated range. The MPJPE drop confirms cross-domain adaptation; the F1 swing shows that fingertip-convention-specific metrics do not transfer across corpora.

\begin{table}[t]
\centering
\small
\begin{tabular}{lcc}
\toprule
Setting & F1 $\uparrow$ & MPJPE (mm) $\downarrow$ \\
\midrule
Tipiano (Für Elise, in-domain) & 0.910 & 32.8 \\
Zero-shot on \ours{}         & 0.93  & 65.9 \\
\;\;+ fine-tune                & 0.66  & 48.8 \\
\bottomrule
\end{tabular}
\caption{Tipiano fine-tuning on \ours{} (5-fold leave-one-pianist-out, professional cohort). In-domain row from~\cite{bae2026tipianocascadedpianohand}.}
\label{tab:tipiano-loso}
\end{table}

\section{Discussion and Conclusion}\label{sec:discussion}\label{sec:conclusion}

We presented \ours{}, a multimodal piano performance dataset pairing real hand and body mocap with audio, MIDI, multi-view video, scores, Visual3D kinematics, and pseudo fingering, structured by technique, difficulty, and expertise. Its fingering pipeline adapts PianoVAM's~\cite{kim2025pianovam} geometry-based scoring to motion capture with $z$-onset refinement and an imputation tier that closes coverage to 100\%, and a Tipiano~\cite{bae2026tipianocascadedpianohand} fine-tuning experiment verifies compatibility with existing motion-from-MIDI pipelines.

The pool of 19 pianists is small, the repertoire is weighted toward technique exercises and Western classical material with no extended techniques, and the fingering labels are pseudo annotations carrying ambiguity flags; future work will grow the pool, widen the audit beyond three pieces, and add extended-technique recordings. Even so, \ours{} fills the gap between precise multimodal piano capture and controlled musical structure, supporting MIR research beyond audio and MIDI alone.

\section{Acknowledgments}
This work was supported by Seoul National University Research Grant in 2021 and YAMAHA Corporation Research Grant.

\section{Ethics Statement}\label{sec:ethics}

The recording sessions were approved by the Institutional Review Board of KAIST (approval nos.\ KH2022-190, KH2023-064, and KH2023-235). All participants provided written informed consent covering audio, MIDI, motion capture, and face-anonymized video. The consent specifies the public-release scope per modality. The full dataset is released through the project page under per-modality license terms reflecting this consent structure.

\bibliography{ISMIRtemplate}

\end{document}